\documentclass[pra,onecolumn]{revtex4}%
\usepackage{amsfonts}
\usepackage{amsmath}
\usepackage{amssymb}
\usepackage{subfigure}
\usepackage{graphicx}%
\providecommand{\U}[1]{\protect\rule{.1in}{.1in}}

\begin{document}
\title{Tunable narrow band source via the strong coupling between optical emitter and nanowire surface plasmons}
\author{J. Yang}
\email{yangjie7898@ecust.edu.cn}
\author{G. W. Lin}
\author{Y. P. Niu}
\author{Y. H. Qi}
\author{F. X. Zhou}
\author{S. Q. Gong}
\email{sqgong@ecust.edu.cn}
\affiliation{Department of Physics, East China University of Science and Technology,
Shanghai 200237, China}

\begin{abstract}
The spectrum width can be narrowed to a certain degree by decreasing the coupling strength for the two-level emitter coupled to the propagating surface plasmon. But the width can not be narrowed any further because of the loss of the photon out of system by spontaneous emission from the emitter. Here we propose a new scheme to construct a narrow-band source via a one-dimensional waveguide coupling with a three-level $\Lambda$-type emitter. It is shown that the reflective spectrum width can be narrowed avoiding the impact of the loss. This approach opens up the possibility of plasmonic ultranarrow single-photon source.
\end{abstract}

\maketitle
\emph{Introduction.--- }
In recent years, the coupling between one-dimensional waveguide and optical emitters has attracted considerable attentions because of their wide application in many aspects\cite{Chang-PRL, Chang-NP, nanolett-Gragting, Chang-PRB, Chang-Nat, spontaneous-OL, OE-Wangququan, PRB-Green, OE-Chen, APL-two, OL-two, NV-PRL, Science2011, OE-NV, NJP-three, Li-PRL, Yang-OE, JPB-two}. The two-level emitter within the waveguide acts as a perfect mirror for the light field at resonance\cite{spontaneous-OL, Li-PRL}. The $\Lambda$-type three-level emitter system($\Lambda$LS) can realize electromagnetically induced transparency (EIT) mechanism to control behavior of the probe photon by applying a classical control light beam in the system\cite{OE-NV, NJP-three}. Similarly, two separated two-level emitters system can also obtain EIT-like transmission spectrum by adjusting the distance between the two emitters\cite{APL-two, OL-two, JPB-two}. As one of important applications of EIT, cavity-linewidth narrowing has shown great potential in many different areas such as laser stabilization\cite{laser-stabilization}, high-resolution spectroscopy\cite{narrowline-highresolution1, narrowline-highresolution2, narrowline-highresolution3}, enhanced light matter interaction, and compressed optical energy\cite{narrowline-interaction}. In this letter, we shall investigate the linewidth narrowing effect in the system of optical emitter coupling the nanowire surface plasmons.

As we have known, in the two-level emitter coupling with SPPs systems(TLS), the linewidth of the reflection peak is proportional to $V^2/v_g$, where $V$ is the coupling strength, and $v_g$ is the group velocity of the photons and can be simplified as the velocity of the light\cite{Chang-NP}. So the reflection spectrum can be narrowed in the weak coupling regime\cite{OL-one}. In such a case, the linewidth can be narrowed to a certain degree for TLS. However if the ratio $\Gamma/\Gamma'$ is further decreased, the narrowed reflective spectrum shall be very weak because that most of the spontaneous emission of emitters is guided into the free space or non-radiative emission. In this letter, we propose a new scheme to construct a narrow-band source via a one-dimensional waveguide coupling with a three-level $\Lambda$-type emitter. By applying a far-detuned classical control light beam, the reflective spectrum demonstrates two dissymmetrical peaks. One of the central frequency of the two peaks locates at $f_c=0$. In this case, the  which equivalent to the case that the two-level emitter within the waveguide acts as a perfect mirror for the propagating single surface plasmon at resonance. So the change of reflective spectrum is the same as that of the single peak in TLS. The other peak occurs at the double-photon resonance, which is the ultra-narrow reflective spectrum we need in this letter. And in this case, the excited state of the emitter can be adiabatically eliminated.

It should be pointed out that the spontaneous emission of emitters can either emit into the guided surface plasmons of the nanowire with rate $\Gamma_{pl}$ or into all other possible channels with respective rates $\Gamma'$\cite{Chang-NP}. In general, $\Gamma'$ includes contributions both from emission into free space and non-radiative emission via ohmic losses in the metal wire. Now let us comment on the above two major contributions in detail. The major source is the imperfect coupling of the emitter to the waveguide. Here we can define the ratio $\Gamma/\Gamma'$ to quantify the coupling strength. Currently, the most advanced experimental setup employs a microwave transmission line side-coupled to a flux qubit. In this experiment, a photon incident from the side was back-scattered with 94\% propability, Translated into the current setting this corresponds to a coupling of $\Gamma/\Gamma'\approx32$\cite{QND-EPL}. In the optical regime, a strongly increased decay rate was demonstrated for single quantum emitters coupled to s surface plasmon polariton mode on a silver nanowire\cite{Chang-Nat}. For a single diamond color center, the ratio $\Gamma/\Gamma'\approx2.6$\cite{QND-EPL}. The second possible channel is the photon absorption in the waveguide, losses are particularly strong in plasmonic systems. So a rapid in- and outcoupling to a dielectric waveguide is inevitabe\cite{Chang-NP, Yang-OE, QND-EPL}. Here we need the system work in the weak coupling regime to obtain the narrowed reflective spectrum, which has greatly decreased the requirement of the experiment. In this letter, we show how to overcome this problem by using $\Lambda$LS to obtain extremely narrow linewidth with enough intensity. And we find that the obvious advantage of this proposal is that the central frequency of narrowed spectrum can be tunable by choosing approximate parameters.

The paper is organized as follows. In Section 2, we shall revisit the problem of photon transportation of TLS under a coherent architecture. Then we shall solve the reflective spectrum of $\Lambda$LS based on the theory of one-photon scattering. The effect of large values of $\Gamma'$ on the linewidth narrowing of TLS is also studied. In section 3, we shall propose the scheme to construct a narrow-band source based on the scheme of $\Lambda$LS, and provide the comparison between the TLS and $\Lambda$LS on the effect of linewidth narrowing. It shows that the $\Lambda$LS can overcome the disadvantage of small intensity in the TLS. Finally, our main conclusions are summarized in Section 4.

\emph {Theoretical model}
 In this section, we firstly review the reflective spectrum for the photon in a quasi-one-dimensional traveling electromagnetic modes coupled to one two-level emitter. The configuration of the system is exhibited in Fig.1(a), and the energy level scheme of the two-level emitter is shown in Fig.1(b). The states $\left\vert g\right\rangle$ and $\left\vert e\right\rangle$ have the energy $\Omega_{g}=0$(the energy origin)and $\Omega_{e}=\omega_1$ respectively. The effect of the loss of the photon out of system is also involved. Here we focuses on the effect of the coupling strength on the linewidth narrowing. For simplicity, we fix the parameter $\Gamma'$ and vary the different parameter $\Gamma$ to quantify the coupling strength $\Gamma/\Gamma'$.
\begin{figure}[htbp]
\centering\includegraphics[width=8cm]{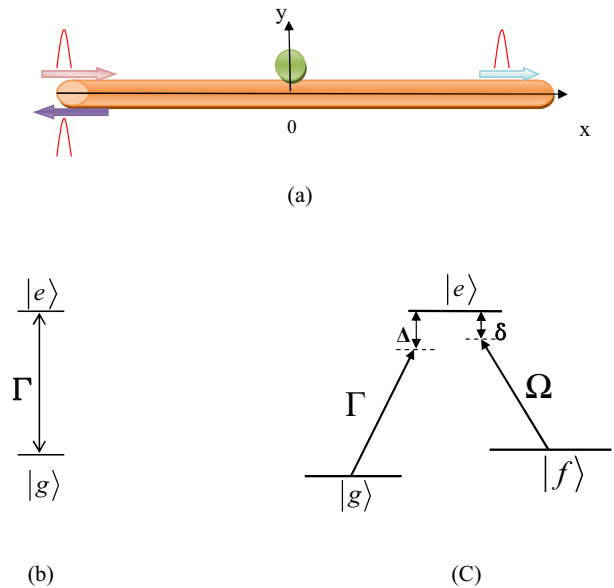} \caption{(color online) (a) Schematic diagram of a metal nanowire couple to a optical emitters. (b) The energy level scheme of two level emitter. (c)The energy level scheme of $\Lambda$-type three level emitter considered in the present paper.}
\end{figure}

The scattering problem for a single photon in a one-dimensional waveguide coupled to a two-level quantum emitter has been solved exactly\cite{Chang-NP, NJP-three, OL-one, PRA-Fan}. Here we assume that the emitter is initially in a ground state and the photon is incident with the energy $E=ck$, then the reflection coefficient for a monochromatic input state with wave number $k$ can be written as
\begin{equation}
r_k=-\frac{1}{1+\Gamma'/\Gamma-2i(\omega_1-ck)/\Gamma}
\text{.}%
\end{equation}

The reflection amplitude is given by
\begin{equation}
R\equiv |r_k|^2
\text{.}%
\end{equation}

We now discuss the change of linewidth as a function of the ratio of $(\Delta-f_{c})/\Gamma'$, where $\Delta=(\Omega_1-ck)$, and $f_{c}$ denotes the central frequency of the reflective spectrum. It should be pointed out that the maximum value of the reflective amplitude for TLS occurs at the resonance, that is to say that the central frequency $f_{c}$ is zero. For comparison, we fix the parameter $\Gamma'$ as 0.1, and the parameter $\Gamma$ is set as 0.1, 0.5, 1 and 5 respectively. The single-photon reflective spectrum in an optical waveguide coupled with an two-level emitter is demonstrated to be a single peak, as shown in Fig.2.
\begin{figure}[htbp]
\centering\includegraphics[width=8cm]{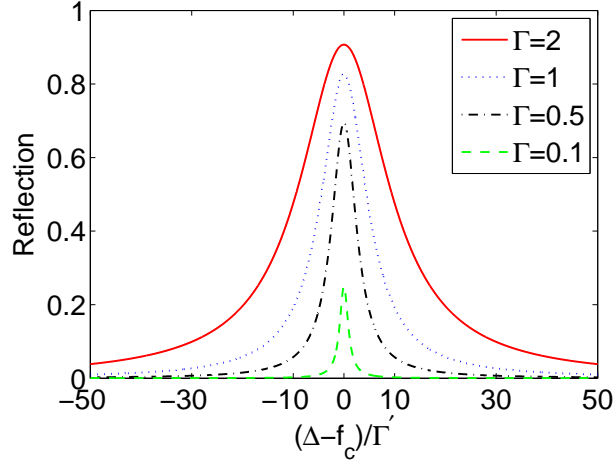} \caption{(color online)The single-photon reflection amplitude as a function of $ (\Delta-f_c)/\gamma_1$ for TLS. The parameter $\Gamma$ is chosen to be 0.1, 0.5, 1 and 5 respectively. The other system parameter $\Gamma'$ is fixed as 0.1. }
\end{figure}

From Fig.2, we shows a typical reflective spectrum for various coupling strength of TLS. At resonance and strong coupling, most of the energy of the plasmon is reflected. It also exhibits that the linewidth of reflective spectrum is getting more and more narrowing with the decrease of the coupling strength. But at the same time, the amplitude of reflective spectrum is also decreased. This is because that most of the photon energy is injected into the free space or decay via ohmic losses in the metal wire when the coupling strength is small. And  for the case of $\Gamma=0.1$, the maximum of the spectrum is only 0.25. Obviously most of the photon energy has been lost. Therefore a new scheme to obtain narrowed spectrum with enough intensity is needed.

In the following, we propose a new scheme to overcome the small intensity of narrowed reflective spectrum by using $\Lambda$LS. The configuration of the system is shown in fig.1(a), and the energy level scheme of $\Lambda$-type three level emitter is shown in fig.1(c).
The nanowire behaves as a 1D continuum for the coherent transport of photons, and the propagating single surface plasmon is described by annihilation operator $a_k$. The states $\left\vert g\right\rangle$ and $\left\vert e\right\rangle$ have the energy $\Omega_{g}=0$(the energy origin)and $\Omega_{e}=\omega_1$ respectively. The propagating single surface plasmon couples to the transition $\left\vert g\right\rangle \leftrightarrow \left\vert e\right\rangle$ with the strength $\Gamma$. The excited emitter's state $\left\vert e\right\rangle$ is coupled to another level $\left\vert f\right\rangle$ by a classical laser beam with Rabi frequency $\Omega$ and detuning $\delta$. In the spirit of the quantum jump picture, spontaneous emission to other nodes out of the one-dimensional waveguide is modeled by attributing an imaginary part $-i\Gamma'$ to the energies of the excited level $\left\vert e\right\rangle$ and an imaginary part $-i\Gamma_2$ to the energies of the metastable level $\left\vert f\right\rangle$ in $\hat{H}_{emitter}$ respectively. We firstly give a general description for the single-photon scattering problem in one dimensional waveguide based on three-level system. The dynamics in such a one-dimensional waveguide is modeled by the total Hamiltonian
\begin{equation}
H_{tot}=H_{free}+H_{emitter}+H_{int}
\text{. }%
\end{equation}
All parts of Eq.(3) can be respectively expressed as
\begin{equation}
H_{free}=
{\displaystyle\int\nolimits}
dx\{-iv_g C_R^+\frac{\partial}{\partial x}C_R+iv_g C_L^+\frac{\partial}{\partial x}C_L\}
\text{, }%
\end{equation}

\begin{equation}
H_{int}= \Gamma(a^{+}_k+a_k)(S_{+}+S_{-})
\text{, }%
\end{equation}

\begin{equation}
H_{emitter}=
(\omega_1-i\Gamma')\left\vert e\right\rangle \left\langle e\right\vert+(\omega_1-\delta-i\Gamma_2)\left\vert f\right\rangle \left\langle f\right\vert+\Omega/2(\left\vert e\right\rangle \left\langle f\right\vert
+\left\vert f\right\rangle \left\langle e\right\vert)
\text{, }%
\end{equation}
where $S_{+}$ and $S_-$ denote the emitter's raising and lowering operators, respectively. In Eq.(4), the parameter $v_g$ is the group velocity of the photons and can be simplified as the velocity of the light\cite{Chang-NP}. The transition between the ground state and the excited state of the emitter is coupled to the propagating surface plasmon modes with the coupling constant $\Gamma$. $\hat{\sigma}_{ee}\equiv\left\vert e\right\rangle \left\langle e\right\vert$ represents the electronic population operator of the excited states. Throughout this paper, we assume that the emitter is initially in a ground state and the photon is incident with the energy $E=ck$, a linear and nondegenerate dispersion relation holds over the relevant frequency range, and a single plasmon is coming from the left with energy $E_k=ck$. Then the stationary state of the system, defined by $H\left\vert \psi_k\right\rangle=E_k\left\vert \psi_k\right\rangle$, can be constructed in the form
\begin{equation}
\left\vert E_k\right\rangle=
{\displaystyle\int\nolimits}
dx[\phi^+_{k,R}(x)C^+_R(x)+\phi^+_{k,L}(x)C^+_L(x)]\left\vert 0,g\right\rangle
\\
+e_{k}\left\vert 0,e\right\rangle+f_{k}\left\vert 0,f\right\rangle
\text{,}%
\end{equation}
where $e_k$($f_k$) is the probability amplitude of the emitter in the excited state $\left\vert e\right\rangle$(metastable state $\left\vert f\right\rangle$), $\left\vert 0,g\right\rangle$ denotes the vacuum state with zero plasmon and the emitter being unexcited, $\left\vert 0,e\right\rangle$ denotes the vacuum field and the emitter in the exited state. $\phi^+_{k,R}$[$\phi^+_{k,L}$] is the wave function of a right-going(a left-going) plasmon, and can be written as

\begin{equation}
\phi^+_L(x)=r e^{-ikx}\theta(-x)
\text{, }%
\end{equation}

\begin{equation}
\phi^+_R(x)=e^{ikx}[t \theta(x)+\theta(-x) ]
\text{, }%
\end{equation}
where $\theta(x)$ is the step function.
By solving the eigenvalue equation $H\left\vert E_k\right\rangle=E_k\left\vert E_k\right\rangle$, we finally get the reflection coefficient as the follows:
\begin{equation}
r_k=-\frac{i \Gamma^2/v_g(\omega_1-E_k-\delta-i\Gamma_2)}{(E_k-\omega_1+i\Gamma')(E_k-\omega_1+\delta+i\Gamma_2)-\Omega^2/4+i\Gamma^2/v_g(E_k-\omega_1+\delta+i\Gamma_2)}
\text{. }%
\end{equation}

The reflection amplitude for the $\Lambda$LS is given by $R= |r_k|^2$.

\section {Tunable narrow band source using $\Lambda$-type three-level emitter system}
In the following, we shall investigate the linewidth narrowing effect of the $\Lambda$LS. First of all, we study the single-photon reflective properties under the resonant control field ($\delta=0$). And our emphasis is still put on the change of linewidth with the coupling strength, as shown in Fig.3. Here the reflective spectrum appears a multiple 'peak-dip-peak' structure (EIT-like reflective spectrum) because of the interference between the two transitions. This is different from a single peak structure in the two-level system. But with the coupling strength decreasing, the two peaks is becoming declining and narrowing. This phenomenon is also appear in the above section. Similarly, we cannot obtain narrow reflective spectrum with the intensive amplitude only by the means of decreasing coupling strength in $\Lambda$LS.

\begin{figure}[htbp]
\centering\includegraphics[width=8cm]{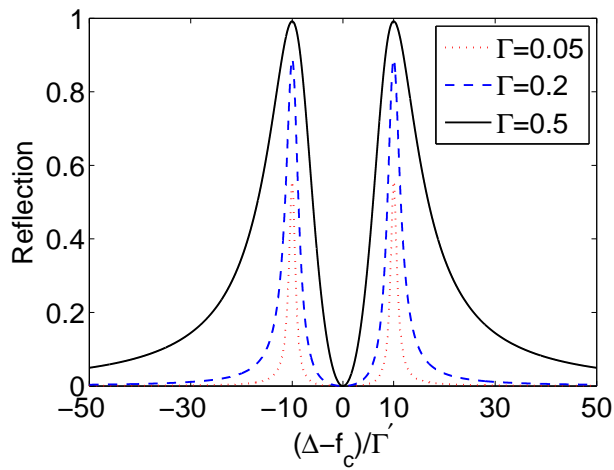} \caption{(color online)The single-photon reflection amplitudes as a function of $(\Delta-f_c)/\Gamma'$ for $\Lambda$LS. The parameter $\Gamma$ is set as 0.05, 0.2, 0.5 respectively. The other parameters are set as follows: $\Omega=2$, $\Gamma'=0.1$, $\Gamma_2=0$, $\delta=0$.}
\end{figure}

Then we adjust the frequency of the control field and study the change of the reflective spectrum under the different detuning $\delta$. As we all know, the weak coupling means that the decay rate to all other channels dominates over the decay rate of the excited atom to the waveguide modes. The obtained reflective spectrum is very faint when we decrease the coupling strength because that most of the spontaneous emission of emitters is guided into the free space or non-radiative emission. Therefore, we should suppress the effects of spontaneous decay of the excited state and decrease the coupling strength at the same time. As we have mentioned above, the transition between the excited state $\left\vert e\right\rangle$ and the metastable state $\left\vert f\right\rangle$ is driven by a classical pulses with the corresponding Rabi frequencies signified by $\Omega$. Here we adopts the method proposed by Law and Eberly\cite{adiabatic1, adiabatic2}. By adjusting this interaction as far-detuned, we find two dissymmetrical peaks. One of central frequency locates at $f_c=0$, whose change is the same as that of the single peak in TLS. The other peak occurs at the double-photon resonance, which is the ultra-narrow reflective spectrum we need in this letter. And in this case, the excited state of the emitter can be adiabatically eliminated. It should be pointed out that the reflective spectrum of $\Lambda$LS has double dissymmetrical peaks and the TLS has only one peak in the reflective spectrum. Here we only use the peak under the case of double-photon resonance to compare with the single peak in TLS. Figure 4 shows the change of linewidth as the function of the ratio of $(\Delta-f_{c})/\Gamma'$. In Fig.4, blue dot line, black dash line and red solid line respectively denote the reflective spectrums of TLS, $\Lambda$LS with $\delta=0$, and $\Lambda$LS with $\delta=5$. All other parameters are chosen as the same except for the detuning $\delta$. From fig.4, we can see that the reflective spectrum of $\Lambda$LS is more intensive than that of TLS under the same coupling strength. For the $\Lambda$LS resonant with the control field, the linewidth of the reflective spectrum has no significant change compared with that of TLS. But when we increase the detuning $\delta$ and keep other parameter as the same, the linewidth can be narrowed greatly and the intensity can be highly preserved.
\begin{figure}[htbp]
\centering\includegraphics[width=8cm]{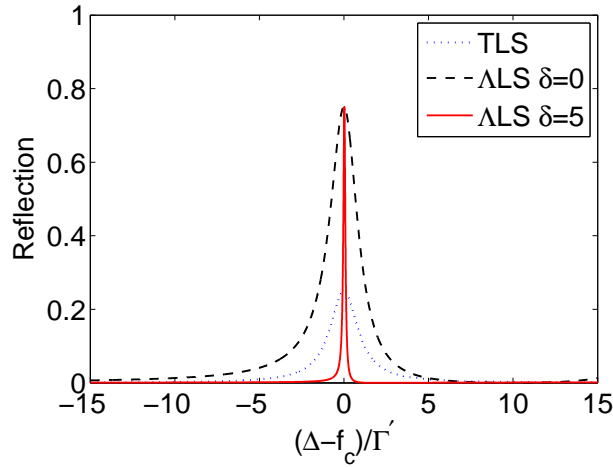} \caption{(color online)The comparison of photon reflection amplitudes as a function of $(\Delta-f_c)/\Gamma'$ for TLS and $\Lambda$LS. For TLS, the parameters are set as follows: $\Gamma=\Gamma'=0.1$. For $\Lambda$LS, the detuning $\delta$ is set to be 0 for the red solid line and 5 for the black dash line respectively. And all other parameters are set as follows: $\Gamma=\Gamma'=0.1$, $\Omega=2$, $\Gamma_2=0$.}
\end{figure}

To see the effect of the detuning on the linewidth narrowing more clearly. In the following, we shall compare the reflective spectrum under the different detuning $\delta$, as shown in Fig.5. We still choose the other parameters as the same besides the detuning $\delta$. Here black dash line, blue dot line and red solid line respectively denote the reflective spectrums of $\Lambda$LS with $\delta=5$, $\Lambda$LS with $\delta=10$, and $\Lambda$LS with $\delta=15$. We can see that the reflective spectrum is getting more and more narrowing with the increase of the detuning $\delta$. But the reflective amplitude is always maintained about 80\%. This property can be utilized to construct a tunable narrow band source. And the needed narrowed frequency and linewidth can both be obtained by choosing approximate parameters.

\begin{figure}[htbp]
\centering\includegraphics[width=8cm]{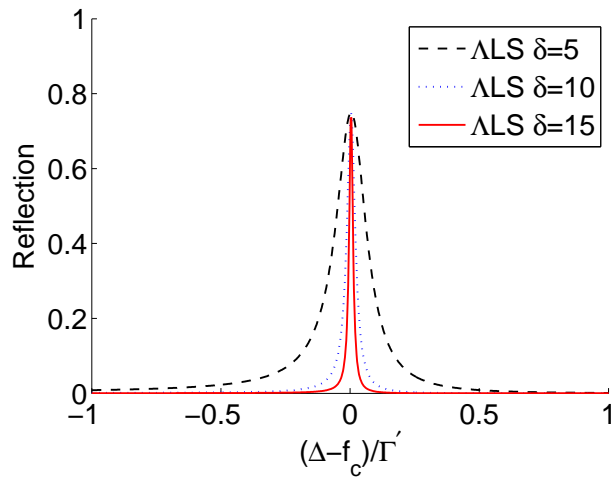} \caption{(color online)The comparison of photon reflection amplitudes as a function of $ \Delta/\gamma_1$ for $\Lambda$LS with different $\delta$. The detuning $\delta$ is set to be 5 for the black dash line, 10 for the blue dot line, and 15 for the red solid line respectively. And all other parameters are set as follows: $\Gamma=\Gamma'=0.1$, $\Omega=2$, $\Gamma_2=0$.}
\end{figure}

\emph{Conclusions}

In conclusion, we have proposed a scheme to construct a narrow-band source via a one-dimensional waveguide coupling with a three-level $\Lambda$-type emitter. Through rigorous theoretical calculation, we find that a ultra-narrow reflective spectrum with intense amplitude can be obtained for the system work in a weak coupling regime and under a far-detuned control field. It is shown that the central frequency of narrowed spectrum can be tunable by choosing approximate parameters. And the system should work in the weak coupling regime to obtain the narrowed reflective spectrum, which has greatly decreased the requirement of the experiment. This approach opens up the possibility of plasmonic ultranarrow band source.

\textbf{Acknowledgments: }The work is supported by the the National Natural Science Foundation of China(11347155, 11274112, 11204080), the Fundamental
Research Funds for the Central Universities(WM1214019), and China Post-doctoral
Science Foundations (Grant Nos.2011M500739).

\end{document}